\documentclass[journal]{IEEEtran_RSIFSMA}
\usepackage[T1]{fontenc} % para hifenização
\usepackage{ae} 
\usepackage[style=ieee]{biblatex}
\newcommand{\RSIFSMA}{Revista de Sistemas de Informação da FSMA}
\newcommand{\NUM}{25}
\newcommand{\ANO}{2020}
\newcommand{\PP}{56-65}
\newcommand{\HFP}{%
\begin{tabular}{p{\columnwidth}p{\columnwidth}}%
\hfill \RSIFSMA \hfill~                                                                       
& \hfill \includegraphics[width=0.5\columnwidth]{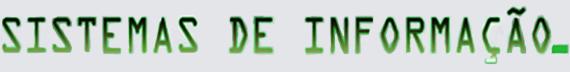} \hfill~  \\%
\hfill n. \NUM\ (\ANO) pp. \PP \hfill~                                               
& \hfill http://www.fsma.edu.br/si/sistemas.html \hfill~ \\ \\%                                                                      
\includegraphics[width=\columnwidth]{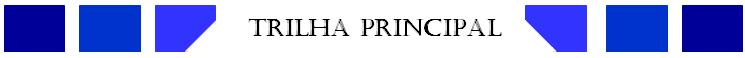} & \\%
\end{tabular}%
}
\newcommand{\TITULO}{Models of Computing as a Service and IoT: an analysis of the current scenario with applications using LPWAN}
 \newcommand{\AUTORES}{BEZERRA, W.R., KOCH, F.L., WESTPHALL, C.B.}
\usepackage[pdftex]{graphicx}
\usepackage{lineno}
\usepackage{color, soul}
\usepackage{multirow}
\usepackage{subfigure}

\begin{document}

\def\figurename{Fig.} 
\renewcommand{\tablename}{Tab.}
\title{~\\[2ex] \TITULO}
%%%%%%%%%%%%%%%%%%%%%%%%%%%%%%%%%%%%%%%%%%%%%%%%%%%%%%%%%%%%%%%%%%%%%%%%%%%%
%%%%%%%%%%%%%%%%%%%%%%%%%%%%%%%%%%%%%%%%%%%%%%%%%%%%%%%%%%%%%%%%%%%%%%%%%%%%


\author{
Wesley dos Reis Bezerra,
Fernando Luiz Koch,
Carlos Becker Westphall,
\thanks{Corresponding author: Wesley dos Reis Bezerra, \textit{wesleybez@gmail.com}}
}

\markboth{\scriptsize \AUTORES\ / \RSIFSMA\ n. \NUM\ (\ANO) pp. \PP}%
{\AUTORES: \TITULO}
%%%%%%%%%%%%%%%%%%%%%%%%%%%%%%%%%%%%%%%%%%%%%%%%%%%%%%%%%%%%%%%%%%%%%%%%%%%%
%%%%%%%%%%%%%%%%%%%%%%%%%%%%%%%%%%%%%%%%%%%%%%%%%%%%%%%%%%%%%%%%%%%%%%%%%%%%

\maketitle

%%%%%%%%%%%%%%%%%%%%%%%%%%%%%%%%%%%%%%%%%%%%%%%%%%%%%%%%%%%%%%%%%%%%%%%%%%%%
%*** RESUMO EM INGLÊS
%%%%%%%%%%%%%%%%%%%%%%%%%%%%%%%%%%%%%%%%%%%%%%%%%%%%%%%%%%%%%%%%%%%%%%%%%%%%
\renewcommand{\abstractname}{Abstract}
\begin{abstract}
\boldmath
%%%%%%%%%%%%%%%%%%%%%%%%%%%%%%%%%%%%%%%%%%%%
%% 00 ABSTRACT
%%%%%%%%%%%%%%%%%%%%%%%%%%%%%%%%%%%%%%%%%%%%

This work provides the basis to understand and select Cloud Computing models applied for the development of IoT solutions using Low-Power Wide Area Network (LPWAN). Cloud Computing paradigm has transformed how the industry implement solution, through the commoditization of shared IT infrastructures. The advent of massive Internet of Things (IoT) and related workloads brings new challenges to this scenario demanding malleable configurations where the resources are distributed closer to data sources. We introduce an analysis of existing solution architectures, along with an illustrative case from where we derive the lessons, challenges, and opportunities of combining these technologies for a new generation of Cloud-native solutions.

\end{abstract}
%%%%%%%%%%%%%%%%%%%%%%%%%%%%%%%%%%%%%%%%%%%%%%%%%%%%%%%%%%%%%%%%%%%%%%%%%%%%

%%%%%%%%%%%%%%%%%%%%%%%%%%%%%%%%%%%%%%%%%%%%%%%%%%%%%%%%%%%%%%%%%%%%%%%%%%%%
%*** PALAVRAS-CHAVE EM INGLÊS, SEPARADAS POR VÍRGULAS
%%%%%%%%%%%%%%%%%%%%%%%%%%%%%%%%%%%%%%%%%%%%%%%%%%%%%%%%%%%%%%%%%%%%%%%%%%%%
\renewcommand{\IEEEkeywordsname}{Index Terms}
%%%%%%%%%%%%%%%%%%%%%%%%%%%%%%%%%%%%%%%%%%%%%%%%%%%%%%%%%%%%%%%%%%%%%%%%%%%%

% \fi
\IEEEpeerreviewmaketitle

%%%%%%%%%%%%%%%%%%%%%%%%%%%%%%%%%%%%%%%%%%%%%%%%%%%%%%%%%%%%%%%%%%%%%%%%%%%%%%%%%%%%%%%%%

%% PART 01 - INTRODUCTION 
%%
\section{Introduction}
\label{sec:intro}
%%%%%%%%%%%%%%%%%%%%%%%%%%%%%%%%%%%%%%%%%%%%
%% 01 INTRODUCTION
%%%%%%%%%%%%%%%%%%%%%%%%%%%%%%%%%%%%%%%%%%%%

%% What is important?
%%
Cloud Computing is the engine to modern Internet of Things (IoT) solution development. However, this paradigm was originally designed for the purpose of shared IT infrastructure, primarily allowing for business of any size to trade on-premise infrastructure by a rented resources \cite{dsouza2014policy, mell2011nist}. 

As workloads start to migrate to IoT-based solutions, there are new challenges around distribution, heterogeneity, volume, velocity, variety, security, vulnerability and others \cite{aazam2014fog, owaspTop10, deep2019survey, bertin2019access}. Hence, there is a need to evolve the Cloud Computing paradigm with malleability, distribution, and closer proximity to the data sources. This is the origin of mixed models like Edge Computing, Fog Computing, Mist Computing and others.

On the other hand, the utilisation of Low-Power Wide Area Networks (LPWAN) and \textit{publish-subscribe protocols} like $AMQP$, $MQTT$, $STOMP$, and $CoAP$, is increasingly popular in Cloud-IoT solution design. New challenges in combining these models revolve around issues of synchronisation, configuration, security, vulnerability and others \cite{owaspTop10,deep2019survey,yi2015fog}. For instance, more sophisticated security mechanisms require larger computing capacity, such as processing, memory utilisation, and power consumption. Engineers must measure the trade-offs between performance, security, and expected device cost while designing secure IoT devices and deploying distributed Cloud Computing configurations. 

%% What is missing
%%

We argue that these issues can be mitigated by selecting the appropriate model combination for the solution demand. Solution designers and application developers need to understand the characteristics and capability of the diverse configurations and how they align with the system requirements in hands. Therefore, our research question is:
\begin{quote}
    What is the best Cloud Computing model to be applied for a given application scenario involving Cloud-IoT-LPWAN?
\end{quote}

%% What are we doing to get there?
%%

We introduce an analysis of the existing models of Cloud Computing applicable for the development of solutions involving IoT, LPWAN and constrained devices. Our goal is to provide the basis for researches and solution designers to compare and select technologies for product development. We acknowledge the limitations of this study as this is an extensive area, permeated by different challenges, opportunities and requirements. Nevertheless, we deem this analysis comprehensive enough to offer the basis for understanding and analysis of the combinations. This study provides three main contributions to the field:

\begin{enumerate}
    \item a survey of Cloud Computing paradigms applicable for this application scenario;
    \item an analysis of the challenges and opportunities in developing Cloud-native applications in this scope;
    \item guidance on selecting the best combination of Cloud Computing model and LPWAN in different application scenarios.
\end{enumerate}

%This work is organised as follows. Section \ref{sec:background} provides an overview of the background scenario. Section \ref{sec:survey} introduces a survey of the Cloud Computing paradigms, support and challenges when applied for the problem scenario. Section \ref{sec:proposal} analysis the support of the different paradigms when applied for an illustrative scenario. We conclude with a discussion of the lessons learned, challenges and opportunities in Section \ref{sec:conclusion}.

%% PART 02 - BACKGROUND 
%%
\section{Background}
\label{sec:background}
%%%%%%%%%%%%%%%%%%%%%%%%%%%%%%%%%%%%%%%%%%%%
%% 02 BACKGROUND
%%%%%%%%%%%%%%%%%%%%%%%%%%%%%%%%%%%%%%%%%%%%

In this section, we present an overview of the current scenario involving Cloud Computing, connected devices, and the challenges and opportunities to combine Cloud-IoT-LPWAN.

%%%%%%%%%%%%%%%%%%%%%%%%%%%%%%%%%%%%%%%%%%%%%%%%%%%%%%
%%%%
\subsection{Cloud Computing}
\label{sec:cc}

Cloud Computing can be described as a a model to support shared Information Technology (IT) resource accessible on-demand through a network infrastructure \cite{mell2011nist}. It encompasses a pool of computational resources, like processing, storage, applications, and other services, which are made available through virtualised environments, accessed through the global network infrastructure. This form of computing is increasingly perceived as the 5th utility (after water, electricity, gas, and telephony), which provides the basic level of computing service that is considered essential to meet the everyday needs of the general community \cite{gubbi2013internet, buyya2010cloud}.

Operationally, Cloud Computing is segmented in three types of services \cite{mell2011nist}: Infrastructure as a Service (IaaS), where the shared resources relate to computing infrastructure, like Virtual Machines, virtual disks, and others; Platform as a Service (PaaS), where the Cloud infrastructure provides virtualised operational platforms, and; Software as a Service (SaaS) where the consumer has access to shared software running on the Cloud. Each service model has its pros and cons and their adoption relates to the business requirements. For instance, the SaaS model has been widely used by software developing companies to provide solutions through \emph{web platform} without the need to maintain the complete stack -- hardware, OS, HTTP servers, and access; the Cloud SaaS provides all-in-one service accessible through the Web, also including maintenance, support, security, reliability, elasticity, scalability, and others. 

The Cloud Computing model introduces a form of distributed computing boosted by a major business prerogative: trade the fixed cost infrastructure around on-premise computing by shared IT infrastructure with variable cost \cite{dsouza2014policy}. The business argument is that Cloud clients do not need to afford computing equity, including machinery and physical infrastructure, but instead rent this service from a Cloud provider who also takes care of maintenance, support, depreciation, Quality of Services, infrastructure, redundancy, safety, security, and others.  Combine with advances in network communication and scalability of computing power, this model is rapidly becoming the \emph{de facto} infrastructure for Digital Transformation strategies \cite{matt2015digital, berman2012digital}.

%%%%%%%%%%%%%%%%%%%%%%%%%%%%%%%%%%%%%%%%%%%%%%%%%%%%%%
%%%%
\subsection{Distributed Cloud Computing}
\label{sec:distributedcc}

However, as business start to migrate workloads to IoT-based solutions, there are many challenges to the Cloud Computing model to support the new computational demands \cite{aazam2014fog, owaspTop10, deep2019survey, bertin2019access}. The term IoT was introduced by Kevin Ashton in 1998 \cite{deep2019survey} and has been used to designate smart devices with internet connectivity. IoT is flourishing as an important tool to solve problems in several areas of knowledge, such as Smart Cities, Smart Buildings, Industry 4.0, Precision Agriculture, Health, Education, Connected Vehicles, and many others \cite{sha2018security, ramli2019toward, aazam2014fog}. Each area of IoT application has its specified demands for network consumption, latency sensitivity, physical network layer, distribution, power consumption, and security. Due to this diversity, a large number of companies have developed IoT solutions leading to a heterogeneous and fragmented market \cite{sanaei2013heterogeneity}.

Hence, there is a need to adapt the Cloud paradigm with malleability, distribution, and closer proximity to the data sources. This is the origin of mixed models like Edge Computing, Fog Computing, Mist Computing and others. There is an increased need for localised processing and storage in distributed IoT solutions. Even though usually IoT data is formed by small packages, due to the large number of devices, they generate large data volumes for communication, storage, and processing \cite{vaquero2014finding}. Therefore, there is a need to select the proper distributed cloud computing  to support the requirements of specific IoT solutions \cite{bonomi2014fog}.

Telemetry protocols have been incorporated into existing Cloud Computing services to address the issues of data volume and velocity in IoT configurations. To cite the main ones, MQTT \cite{standard2014mqtt}, CoAP \cite{shelby2014constrained}, DDS \cite{pardo2005introduction}, AMQP \cite{standard2012oasis}, XMPP \cite{saint2011extensible}, STOMP \cite{stompRfc} and HTTP19 are example of popular publish-subscribe protocols New challenges in combining these models revolve around issues of synchronisation, configuration, security, vulnerability and others \cite{owaspTop10,deep2019survey,yi2015fog}. For instance, more sophisticated security mechanisms require larger computing capacity, such as processing, memory utilisation, and power consumption. Engineers must measure the trade-offs between performance, security, and expected device cost while designing secure IoT devices and deploying distributed Cloud Computing configurations.

%%%%%%%%%%%%%%%%%%%%%%%%%%%%%%%%%%%%%%%%%%%%%%%%%%%%%%
%%%%
\subsection{Challenges and Opportunities}
\label{sec:gap}

With the IoT bringing so many connected devices, some concerns come to the forefront, such as aspects of:
\begin{itemize}
    \item security, where flaws are frequently reported by data manufacturers \cite{owaspTop10, deep2019survey, bertin2019access}, system intrusion  \cite{schulter2006grid, schulter2006towards, vieira2019autonomic}, and others;
    \item \textit{data volume and data flows} pose a problem to IoT applications, where it is important to understand the characteristics of the data flow between: (i) \textit{small and simple data streams}, e.g. coming from e.g. smart metering applications \cite{chowdhury2020adaptive}, and;  (ii) \textit{complex and larger data flows }, e.g. sources from images, patterns and matrices \cite{muhammad2019cost};
    \item \textit{limited or heterogeneous networks};
    \item \textit{energy consumption}, which becomes a  major challenge when applied to constrained devices \cite {girgenti2019feasibility}, often designed to perform with a long battery life, e.g. ultra-long life battery can last for 10 years if the device is properly operated.
\end{itemize}

Finally, there is a surge of implementations around \emph{Low Power Wide Area Network} (LPWAN), where devices connect to centralised services through wireless protocols with limited data transmission and radio link.

%% PART 03 - SURVEY 
%%
\section{Configurations}
\label{sec:survey}
%%%%%%%%%%%%%%%%%%%%%%%%%%%%%%%%%%%%%%%%%%%%
%% 03 SURVEY
%%%%%%%%%%%%%%%%%%%%%%%%%%%%%%%%%%%%%%%%%%%%

As IoT-based solutions start to take hold of the market, it became clear that Cloud-centred solutions imposed severe restrictions in terms of latency, heterogeneity, volume, velocity, variety, security, vulnerability and others. For that, new concepts of distributed Cloud Computing started to emerge in the market, each bringing new solutions and challenges of their own. 

Computing as a service has gone through several stages of development. It can be used in virtual automation through Virtual Machine Environment (VME), the virtualisation of services in Software Define Networks (SDN), and the distributed models like Fog Computing, Mist Computing, Mobile Edge Computing, Mobile Cloud Computing, and Superfluid Computing, between others.

In what follows, we introduce these alternative models and explain how they support application scenarios involving Cloud-IoT-LPWAN. We also interwove an analysis of the challenges and opportunities in developing Cloud-native applications in this scope.

%% SUB-PARTS ARE LINKED BY MAIN.TEX

\subsection{Cloud Computing-IoT}
\label{sec:cc-iot}
%%%%%%%%%%%%%%%%%%%%%%%%%%%%%%%%%%%%%%%%%%%%
%% 03 SUBPART - CLOUD CENTRED
%%%%%%%%%%%%%%%%%%%%%%%%%%%%%%%%%%%%%%%%%%%%

\begin{figure}[h!]
    \centering
    \includegraphics[width=0.85\linewidth]{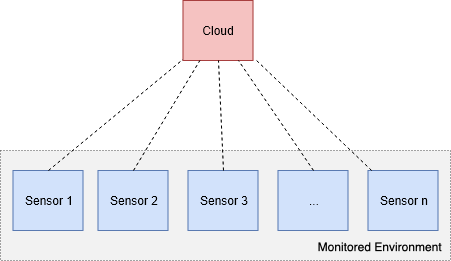}
    \caption{Cloud Computing-Centred IoT Architecture}
    \label{fig:cc_iot_arch}
\end{figure}

There are some case where Cloud Computing looks like the ideal option for IoT. This model provides low cost processing and storage, availability, well-known programming resources, and others The concept works well in situations where wither there is good connectivity or no demand for real-time processing. Popular examples are applications of video security in home automation that store streams on the Cloud. Li et al \cite{li2018security} presents the example of Connected Vehicles Cloud Computing where vehicle-bounded sensor devices connect to Cloud-centred infrastructure for services. Figure \ref{fig:cc_iot_arch} depicts the solution architecture, composed of:

\begin{enumerate}
    \item \emph{cloud layer}, providing centralised processing and storage; 
    \item \emph{service consumers layer}, which in the case of IoT applications service consumers are sensors.
\end{enumerate}

Sensors must be connected either directly through Application Program Interfaces (APIs) or through a Protocol Gateway. All services are performed on the Cloud infrastructure, including processing, storage, and any other add-value service. If location information is needed, the gateway (or sensor) must include location information such as pre-configured or from a GPS, as the central service lacks awareness of location \cite{luan2015fog}.

This architecture presents challenges to support IoT-based applications mainly due to volume, latency, heterogeneity, security, and others. Basu et al \cite{basu2018cloud} list some important challenges, such as: data segregation, data location, data incomplete data, monitoring and data logging, problems associated with the security of Virtual Machines and their environment, and even natural disasters. Subramanian et al \cite{subramanian2018recent} enlist confidentiality and integrity  among the main security challenges. 

Other requirements are authentication, auditing, and legal security requirements. The latter is important because the data is subject to regulations in some countries. Sha et al \cite{sha2018security} describe the challenges around heterogeneous network technologies, privacy, large scale of systems, and management of trust. Stergiou et al \cite{stergiou2018security} corroborate to the narrative, listing key challenges around heterogeneity, performance, reliability, big data and monitoring.

\subsection{Fog Computing}
\label{sec:fogcomputing}
%%%%%%%%%%%%%%%%%%%%%%%%%%%%%%%%%%%%%%%%%%%%
%% 03 SUBPART - FOG COMPUTING 
%%%%%%%%%%%%%%%%%%%%%%%%%%%%%%%%%%%%%%%%%%%%

\begin{figure}[ht]
    \centering
    \includegraphics[width=0.85\linewidth]{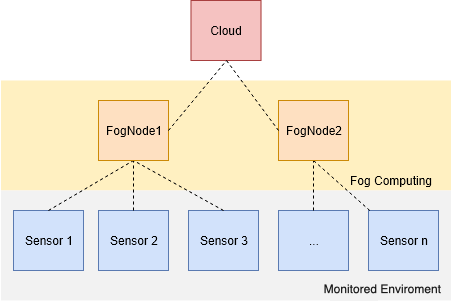}
    \caption{Fog Computing Architecture}
    \label{fig:fc_arch}
\end{figure}

\emph{Fog Computing} is a model of distributed Cloud Computing designed to cope with the growth of IoT environment and issues of latency inherent to this configuration \cite{buyya2019fog, luan2015fog}. Yi et al \cite{yi2015fog} postulate that this model will be the internet of the future due to the number of existing devices and the arrival of more devices in the market.

This model is organised in three layers, as depicted in Figure \ref{fig:fc_arch}:
\begin{enumerate}
    \item Cloud with the classic CC model and centralised computing;
    \item Fog Nodes are central part of the architecture; these can be either physical components or virtual components and are tightly coupled with smart devices or network access nodes, providing computational resources to these devices.
    \item Sensor Nodes providing the data sources in the edge layer.
\end{enumerate}

This model is inherently distributed geographically, providing resources near the data sources where they are usually most necessary. For example, in case of IoT and wearable devices, impeding restrictions of processing and storage in these devices demand for Fog Nodes to deliver situated resources. 

However, the multi-layered nature of this model increases its implementation complexity. One direct impact of this extra complexity is on  security. Yi et al \cite{yi2015fog} describes the challenges of Fog Computing around the choice of virtualisation technologies to be applied, along with the issues of latency, network management, security, and information privacy. Luan et al \cite{luan2015fog} includes challenges around application development of applications, scalability, and distribution. 

DSouza et al \cite{dsouza2014policy} mentions that Fog Computing's multi-level collaboration brings a new set of problems around identity management, resource access management, distributed decision, dynamic load balancing, quality of service and security. Vaquero et al \cite{vaquero2014finding} mentions that in order for Fog Computing to become reality, developers must first solve its many challenges around synchronisation, device discovery, management, security, standardisation, accountability, billing of mobile applications.

\subsection{Mist Computing}
\label{sec:mistcomputing}
%%%%%%%%%%%%%%%%%%%%%%%%%%%%%%%%%%%%%%%%%%%%
%% 03 SUBPART - MIST COMPUTING 
%%%%%%%%%%%%%%%%%%%%%%%%%%%%%%%%%%%%%%%%%%%%

\begin{figure}
    \centering
    \includegraphics[width=0.60\linewidth]{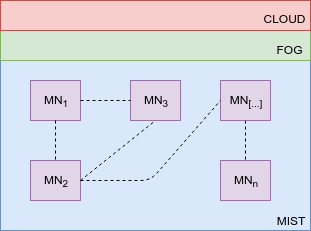}
    \caption{Mist Computing Architecture}
    \label{fig:mc_arch}
\end{figure}

\emph{Mist Computing} (MC) plays an important role in the migrating computation power to the systems' edge. This model extents \emph{Fog Computing} by adding an extra layer placed closer to the edge, below the Fog Nodes \cite{iorga2018fog}. Figure \ref{fig:mc_arch} depicts the architecture:

\begin{enumerate}
    \item \emph{Cloud Computing}  provide centralised services, when needed, such as monitoring, updates, central data repository, offloading processing, and others.
    \item \emph{Fog Computing} provides distributed capabilities and group control over the mist nodes, bridging with Cloud Computing services when required; for instance, \emph{Fog Nodes} may provide resources like regional storage, processing offloading, monitoring, software updates, and others.
    \item \emph{Mist Nodes} are located at the border, providing local resources and processing, with connectivity with their peers and requesting resources from them.
\end{enumerate}

\emph{Mist Nodes} are usually implemented through devices that offer basic computational power, such as Arduinos, Nodemcus, and other microcontrollers and microchips. A practical example of Mist Computing implementation exist in automated vehicles where the multiple elements are connected to a central car unit (Fog), which can connected to other cars for collaboration and collective intelligence. 

The key concept is to promote interaction between \emph{Mist Nodes} as much as possible, refraining from utilising centralised services or devices. The architecture can be conceptualised as a model in which network edge devices have predictable accessibility and provide their communication and computational resources as a service \cite{liyanage2016mepaas}. Mist Nodes can distribute software processes to run on service providers on their own. That is, Mist Computing favours a model of computing ``a hop away'' \cite{vasconcelos2019cloud}.

Mist Computing brings computing power deeper into the edge, embedding processing in microcontrollers and System on Chips \cite{swain2019rise}. This model can provide a flexible environment for execution of customised programs \cite{liyanage2016mepaas}. However, Mist Nodes do not have the computational power of a Fog Node so they must be used in complement. In situations where applications demand processing or storage, these requests can be offloaded to Fog Nodes, which in turn can relay to Cloud Computing. 

This architect can be leveraged to extend the capacity of constrained devices, like mobile sensors, and promote fast processing at the very edge of the system. Being an integral part of the edge, Mist Computing provides the lowest latency possible for an IoT application. However, it suitable only for a restricted number of application scenarios considering the constrained computing capabilities and implicit distribution.

Preden et al \cite{preden2015benefits} describes the challenges of \emph{Mist Computing} related to communication complexity and self-management, as a result of the dependence of a central component. For Yogi et al \cite{yogi2017mist} list as challenges:  low storage capacity, limited bandwidth, and resistance from solution developers to adopt the model. Vasconcelos et al \cite{vasconcelos2018bio} argues that the challenges are common to Fog Computing, and the most characteristic challenge is related to the complexity brought by the dynamic topology. Suarez-Albela et al.\cite{suarez2018practical} list challenges related to security, as the use of security mechanisms such as encryption demand a require energy consumption by the devices.     

\subsection{Mobile Cloud Computing}
\label{sec:mobilecloudcomputing}
%%%%%%%%%%%%%%%%%%%%%%%%%%%%%%%%%%%%%%%%%%%%
%% 03 SUBPART - MOBILE CLOUD COMPUTING 
%%%%%%%%%%%%%%%%%%%%%%%%%%%%%%%%%%%%%%%%%%%%

\begin{figure}[h!]
    \centering
    \includegraphics[width=0.85\linewidth]{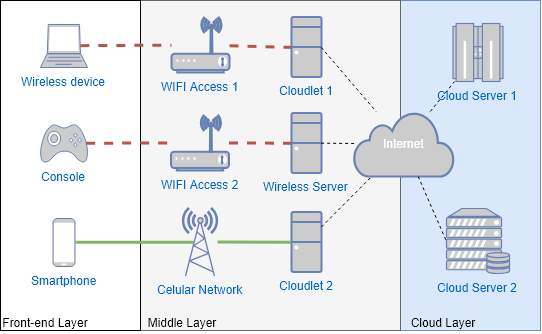}
    \caption{Mobile Cloud Computing Architecture}
    \label{fig:mcc_arch}
\end{figure}{}

\emph{Mobile Cloud Computing} (MCC) is about the provisioning of data and storage closed to the mobile user. In its most elementary form it refers to an infrastructure where both data processing and storage take place outside the mobile device \cite {yi2015fog}. The concept of mobile cloud relates to pervasive devices sharing their support services and heterogeneous resources, such as network band, processing, content, and others \cite{stojmenovic2014fog}. The architecture contains three layers as depicted in Figure \ref{fig:mcc_arch}:

\begin{enumerate}
    \item \emph{Front-end layer}, where either user interfaces or services request applications are located, usually running on mobile devices; this layer works by requesting service from external providers, through offloading processing and storage.
    
    \item \emph{Middle layer}, promotes offloading access to computing power hosted on servers, wireless network, or \emph{cloudlet} \cite{zhang2015offloading}.
    
    \item \emph{Cloud layer}, provides the back-end infrastructure to respond to all requests.
\end{enumerate}

\emph{Mobile Cloud Applications} can transport computing, energy and data storage out of phones and into the Cloud. This allows to create a new range of solutions not only for mobile phone applications but also a range of other solution niches \cite{dinh2013survey}.

In order to prevent delays in processing the data due to latency or network bottlenecks, the network segment between the \textit{front-end} and the \textit{middle layer} must support high throughput. Moreover, the communication link between the \textit{middle layer} and the \textit{cloud} is usually the Internet, thus subject to intermittent connections, latency, and security issues. Hence, real-time applications must be processed in the \textit{middle layer}. The element responsible for bringing \textit{cloud} capabilities close to the mobile device are the \textit{cloudlet} \cite{stojmenovic2014fog}.

Au et al \cite{au2018privacy} argues that the key challenges for this model are related to data and authentication, including: authentication of mobile users and devices, security in data communication and storage, data integrity, data search, and secure data sharing. Leppanen and Riekki \cite{leppanen2019energy} enlist as challenges offloading, scheduling, monitoring, resource tracking, context awareness, and remote service availability. Challenges and issues of heterogeneity in Mobile Cloud Computing are largerly discussed in \cite{sanaei2013heterogeneity}.

Sekaran, Vikram and Chowdary \cite{sekaran2019design} present the issue of security and Distributed Denial of Services (DDoS), and describe different ways to prevent these attacks in MCC. Noor et al \cite{noor2018mobile} list as the main challenges security, privacy, bandwidth control, data transfer, data management, synchronisation, energy efficiency and heterogeneity.

\subsection{Mobile Edge Computing}
\label{sec:mobileedgecomputing}
%%%%%%%%%%%%%%%%%%%%%%%%%%%%%%%%%%%%%%%%%%%%
%% 03 SUBPART - MOBILE EDGE COMPUTING 
%%%%%%%%%%%%%%%%%%%%%%%%%%%%%%%%%%%%%%%%%%%%

\begin{figure}[ht]
    \centering
    \includegraphics[width=0.85\linewidth]{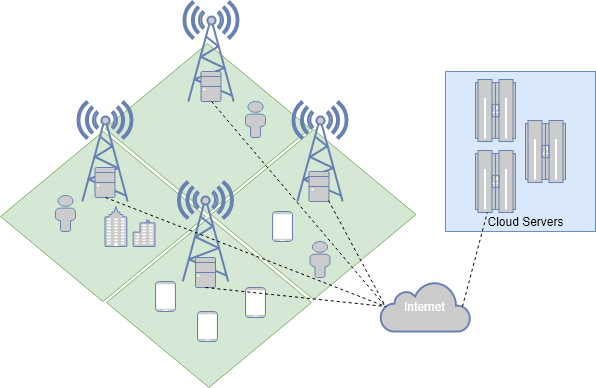}
    \caption{Mobile Edge Computing Architecture \cite{salman2015edge}}
    \label{fig:mec_arch}
\end{figure}{}

Mobile Edge Computing (MEC) utilises the infrastructure based on mobile networks to provide connectivity to edge devices. This architecture has the purpose of bringing computing resources, mainly processing and storage, near to the data sources. Services are deployed directly at base stations or in smart cells, as for instance Femtocells, Picocells, Nanocells, and others. This approach relies on the infrastructure provided by mobile phone operators and it may incur in issues of coverage, high costs of mobile data, and others. 

The architecture is implemented in three layers, as depicted in Figure \ref{fig:mec_arch}:
\begin{enumerate}
    \item \textit{edge}, usually composed of IoT , mobile computing, and wearable devices;
    \item \textit{servers}, providing located computing resources aiming at low latency and quick response for service requests;
    \item \textit{cloud}, providing centralised processing and storage.
\end{enumerate}  

MEC brings some relevant benefits such as ultra-low latency, high bandwidth, real-time access to radio network information, and location services  \cite{zhang2014joint}. For the latter, unlike other models, MEC allows to identify the location of the data source, thus providing additional support for security, locality, and auditing e.g. in case of regulated installations. Moreover, ultra-low latency favours applications that require quick response in decision-making.

The devices located on the edge of the MEC must have components that allow them to access the cellular network, either via \emph{macrocells} i.e. antennas and other classic devices or \emph{smartcells}, which are eager to increase the reach of cellular networks, by expanding their capacity to provide connectivity, specially to o rural areas.

This model is becoming a hot topic with the multiplication of MEC service providers and the arrival 5G technology, in solutions to support communication, computing, control, and content delivery \cite{mao2017survey}. One the key benefits of this model for IoT solutions is better coverage, specially in urban areas. This will allow for like Smart Cities and Smart Buildings that can directly benefit from MEC and 5G. 

The \emph{server layer} brings mobile users close to the benefits of Cloud Computing around elasticity and virtualisation. In the MEC model, this is implemented upon virtualisation platforms like  Network Functions Virtualisation (NFV), Information-Centric Networks (ICN) and Software Defined Networks (SDN) \cite{mao2017survey}. 

NFV promotes the virtualisation of network functions, making them work like in the Cloud Computing model. This is implemented either through dedicated hardware or off-the-shelf devices \cite{zhang2014joint}. The objective of virtualising network functions is towards the cost benefits of shared infrastructure, preventing e.g. capital costs and operational costs related to site allocation, cooling, maintenance, and others. The functioning is based on three concepts: virtualisation, orchestration, and turning all functions into software. Common examples of virtualised functions include firewall, DHCP, NAT, and others

SDN virtualises the network by decoupling control plane from data plane \cite{zhang2014joint}. As an analogy,  NFV visualises network functions, SDN. The goal is to understand the allocation and performance of routes, define new ones, connect physical and virtual services, define policies, allocate IPs, allocate bandwidth, and ensure connectivity. SDN facilities the process of creating NFV by supporting the configuration of  the connection of functions in the network.

Ahmed and Rehmani \cite{ahmed2017mobile} list as challenges the fast development of services at a cost efficient rate, optimised resource utilisation, facilitate the migration of existing application, and security. Vassilakis \cite{vassilakis2016security} mentions the need for security and privacy solutions specific to this model and the possible coexistence with unreliable nodes. Beck et al. \cite{beck2014taxonomy, beck2014mobile} defines some metrics that need to be met to allow utilisation in some areas, such as energy consumption, delay, bandwidth and scalability. They also mention challenges around scalability, mobility awareness, and utilisation awareness by embedded applications.

Varghese et al. \cite{varghese2016challenges} presents a list with five challenges: (i) ability to provide general purpose computing at the edge nodes; (ii) node discovery; (iii) task partitioning and orchestration; (iv) balance demands, processing, and Quality of Service (QoS) and Quality of Experience(QoE); and (v) safety. Roman, Lopes and Mambo \cite{roman2018mobile} list as challenges: infrastructure, virtualisation, resources and tasks, distribution, mobility and programmability.

\subsection{Emerging Models}
\label{sec:emergingmodels}
%%%%%%%%%%%%%%%%%%%%%%%%%%%%%%%%%%%%%%%%%%%%
%% 03 SUBPART - OTHERS
%%%%%%%%%%%%%%%%%%%%%%%%%%%%%%%%%%%%%%%%%%%%

There are two trendy technologies emerging in the market that are worth mentioning at this point: \textit{Superfluid Cloud} and \textit{Cloud Radio Access Network}. We acknowledge that this list is far from exhaustive and other commercial and research solutions emerge frequently in the market, but included a brief description for the reference.

\textit{Superfluid Cloud} \cite{manco2015case} is a multi-tenant model where virtualised services based on software execute on commodity hardware, being shared and deployed across the network. The idea is to apply low-cost devices with devices with significant computational power, such as System-on-Chip, Cubieboard, Raspberry Pi, and others. The most important characteristics include \cite{zhang2014joint}: recursion; scalability; separation between state and processing; support for very small VMs; support for Extended State Finite Machines (XFSMs), and; on-the-fly monitoring.

\textit{Cloud Radio Access Networks} virtualise some important functions of modern telecommunications architecture, lowering the cost of deploying and operating  mobile networks \cite{peng2015fronthaul}. Demestichas et al \cite{demestichas20135g} list challenges around multiple perspectives of society, economy, users, and operators. There are also issues on normalisation in order to have a cohesive, inclusive, and sustainable structure. These challenges relate to wireless communications in general and can be mitigated through C-RAN's centralised structures.

%% PART 04 - PROPOSAL
%%
\section{Use Cases}
\label{sec:proposal}
%%%%%%%%%%%%%%%%%%%%%%%%%%%%%%%%%%%%%%%%%%%%
%% 04 PROPOSAL
%%%%%%%%%%%%%%%%%%%%%%%%%%%%%%%%%%%%%%%%%%%%

In business-critical application that supports decision-making in near real-time based on information from IoT sensors, Quality of Service (QoS) can be measured, between other factors, on aspects of efficiency, accuracy, and security. This sort of solution cannot compromising QoS due to e.g. network latency, intermittent communication, intrusion attacks and things of the sort. It is intuitive that the system will perform better if the decision-making processes are closer to actuators and connected through a reliable communication network, which obviously imply in infrastructure costs. Hence, solution designers juggle to reach a balance between QoS and affordability. 

Let us consider an application scenario in the field of Smart Agriculture:  a system for monitoring artificial tanks in fish farming. In this solution, sensors information about the fish tanks, such as capture water temperature, oxygenation level, ammonia rates, movement, and other information about fishery. The information is relayed to a processing unit for storage, analysis, and recommendation generation. The recommendation system controls actuators that act upon mechanisms to control conditions, e.g. turning the oxygenation pumps, releasing food, promoting water circulation, and providing information and insights to oversees through dashboards on mobile devices. The system also provides visualisation through mobile computing devices to support on-spot decision making. 

This is a business-critical solution as QoS issues on the monitor-actuator system directly impact farm’s production. For instance, taking too long to open oxygenation pumps will lead to high fish mortality; not releasing enough food will lead to malnutrition; providing too much food leads to food poisoning, and others.  Scenarios like this present clear challenges in terms of volume, latency, and intermittent access to remote services even by making use of telemetry protocols, such as MQTT. Hence, solution design around distributed Cloud Computing must be considered.

By applying the \emph{Cloud Computing-IoT model}, described in Section \ref{sec:cc-iot}, developers have the advantage of having centralised data from multiple points, thus supporting insight models that demand correlation of large volumes of data, creation of dynamic filters, and provide greater computational and storage resources. Nonetheless, this model stress the issues of volume, latency, and intermittent connections around the uploading data links. Hence, this model is useful for specific requirements, such as the creation of models that correlate data from multiple sites, and in specific conditions, e.g. when there is a high-throughput stable data link to support the upload data link. 

On the other hand, the \emph{Fog Computing model}, introduced in Section \ref{sec:fogcomputing}, allows for data being stored in an intermediate node, preventing good part of latency for data uploading. The system's intelligence would be allocated in the FN, closer to the edge. Thus, insights, reports, dashboards, and recommendations would be processed on the Fog Nodes, based on Analytic models trained on the Cloud. The process of synchronisation, maintenance, and model revision and retraining,  and retraining, make part of the Orchestration Strategy and must be defined by developers during the system setup. This extra step of complexity is the down side of the model. It requires that developers are familiar of the new demands and toolkits to enable the development at an acceptable cost. Moreover, the data is more expose at the edges, thus highlighting issues of security and privacy.

\emph{Mist Computing}, presented in Section \ref{sec:mistcomputing}, is not applicable to this scenario as this model does not provide support to reporting, storing data, and massive data processing. This model, if directly applicable, would provide the best response time in relation to the sensors and actuators, as processing takes place in the same network segment where the sensors are positioned. Due to the limited processing capability, a combination of \emph{Mist Computing} could be adopted in the architecture to provide reactive responses for alarming systems, for instance.

\emph{Mobile Cloud Computing}, introduced in Section \ref{sec:mobilecloudcomputing},  allows for low-cost sensors to be integrated in the solution. This model requires the implementation of \emph{cloudlet} on the same local wireless network. However, due to extensive data distribution, the compilation and access to reports would depend on an auxiliary system installed on a Cloud Computing system and it would require a synchronisation process to upload data from the cloudlets. This setup implies in higher implementation costs, eventual latency on generating the reports, and maintenance complexity.

\emph{Mobile Edge Computing}, explained in Section \ref{sec:mobileedgecomputing}, provides us with a scenario very close to CF. However, it is applicable when there is no possibility of using a cheaper wireless link, with the cellular network option being the only viable one. This can happen due to the sensed tank being remotely distant from a wifi access point but still covered by the mobile network. It is observed that transmission over the mobile network requires greater use of energy and, consequently, shorter battery life.

When considering the utilisation of LP-WAN in the scenario, this technology favors the use of Fog Computing model. That is, when the solution demands higher power networks such as WiFi and wired network, one could not apply the\textit{cloudlets}. If applying a centralised node , such as in the \emph{Cloud Computing-IoT} model, the solution would require a gateway to bridge between the LPWAN segment and the Internet. If the solution involves a number of constrained devices, then any model that require large message exchange, such as Cloud Computing and Mobile Cloud Computing are not the best choice for the scenario.

Hence, we conclude that Fog Computing model is the intermediate solution for this problem scenario. By providing proximity to the edge for the data source, it allows for lower latency  and fastest response time suitable in mission critical situations. Mobile Edge Computing also prevents itself as a viable solution in this scenario, due to similarities with Fog Computing. However, the approach implies in higher hardware costs.

%% PART 05 - CONCLUSIONS
%%
\section{Conclusions}
\label{sec:conclusion}
%%%%%%%%%%%%%%%%%%%%%%%%%%%%%%%%%%%%%%%%%%%%
%% 05 CONCLUSION
%%%%%%%%%%%%%%%%%%%%%%%%%%%%%%%%%%%%%%%%%%%%

This work brought an overview of emerging paradigms of distribute cloud computing, describing their architecture, applicability and models. We focused on solutions around LPWAN and constrained devices and related the restrictions and characteristics of both models. 

We concluded that Fog Computing is the most adequate paradigm for the proposed scenario, providing the desired features of distribution, orchestration, and normalised interfaces. However, the Mobile Edge Computing model provides similar characteristics with appealing cost structure. Thus, both models must be considered when designing IoT solutions that demand low latency, local processing, high throughput and other related characteristics.

Moreover, we assessed that aspects of security are fundamental in distributed Cloud Computing. We highlight the often overseen issues of \emph{intrusion attacks}. As any other distributed environment, Fog Computing, Mist Computing, MEC, and others are prone to intrusion attacks due to their distribution and heterogeneity nature. It requires methods for distributed intrusion detection and reaction allowing for real-time security and intrusion prevention \cite{schulter2006grid, schulter2006towards}. Vieira et al \cite{vieira2019autonomic} introduces a method to apply Big Data for fast intrusion detection and reaction, claiming that the longer it takes to react to intrusion attacks, the more likely are them to succeed. We believe that these solutions will become increasingly more relevant with the widespread of IoT and distributed Cloud Computing. 

As future work, we propose the use of network simulators to evaluate the protocols to be used together with the paradigms. Simulations could also be applied to evaluate message protocols and security issues. This will allow you to assess what network protocol is most suitable for LPWAN and CD in the proposed scene. In addition, there are potential advancements in distributed processing and swarm computing to be considered and integrated in the models. For example, in Assuncao et al \cite{assuncao2004grids} we introduced a view of \emph{grids of agents} as a implementations were very distributed and interconnected acting elements would implement required services. The raise of Fog Computing and Mist Computing catalyse the need for that kind of infrastructure, creating an opportunity for future research and development. Finally, situation aware  solutions such as presented in \cite{deAssuncao2017content}, a context-aware content delivery implementation, will demand for extended coordination and orchestration in Fog Computing and Mobile Edge Computing, calling for research and development in context-aware orchestration in distributed cloud computing environments.

\ifCLASSOPTIONcaptionsoff
  \newpage
\fi



%\bibliographystyle{IEEEtran}
%\bibliography{bibs/geral.bib,bibs/desafios.bib,bibs/paradigmas.bib}
%\begin{thebibliography}{1}
%\end{thebibliography}

\printbibliography

\end{document}